\documentclass[aps,pre,twocolumn,10pt]{revtex4-1}

\usepackage{graphicx,amsmath,amssymb,bbm}
\usepackage[utf8x]{inputenc}
\usepackage{CJK} 

\begin{document}
\begin{CJK*}{UTF8}{min}
\CJKfamily{gbsn}

\title{Absence of Anderson localization in certain random lattices}

\author{Wonjun Choi}
\author{Cheng Yin (澄 殷)}
\author{Ian R. Hooper}
\author{William L. Barnes}
\author{Jacopo Bertolotti}
\affiliation{Physics and Astronomy Department, University of Exeter, Stocker Road, Exeter EX4 4QL, UK}

\begin{abstract}
We report on the transition between an Anderson localized regime and a conductive regime in a 1D scattering system with correlated disorder. We show experimentally that when long-range correlations are introduced, in the form of a power-law spectral density with power larger than 2, the localization length becomes much bigger than the sample size and the transmission peaks typical of an Anderson localized system merge into a pass band. As other forms of long-range correlations are known to have the opposite effect, i.e. to enhance localization, our results show that care is needed when discussing the effects of correlations, as different kinds of long-range correlations can give rise to very different behavior.
\end{abstract}

\maketitle
\end{CJK*}

Wave transport in multiply scattering media is a complex phenomenon. If the scattering is weak enough the interference effects can be neglected and the wave transport can be described in terms of a diffusion equation~\cite{akkermansbook, sheng}. As the scattering strength increases interference effects reduce the diffusion coefficient, an effect known as weak localization~\cite{bsc}. Once the scattering strength overcomes a certain threshold the diffusion coefficient goes to zero, the system becomes Anderson localized, and no macroscopic transport is possible~\cite{AL, AL2}.

Anderson localization is quintessentially an interference effect that can occur for any kind of wave and thus, although it was originally proposed for electrons~\cite{50yAL}, it has been observed for mechanical waves~\cite{ALac, ALpage}, Bose-Einstein condensates~\cite{ALbec} and electromagnetic waves~\cite{ALem}. It is well understood that the dimensionality of the system plays a major role when it comes to Anderson localization. For 3D systems, when the disorder increases, there is a phase transition between a conductive phase, where all the eigenmodes are extended, and an insulating phase, where the eigenmodes become exponentially localized in regions of size $\sim \xi$ (the localization length)~\cite{ALtheory}.

For 1D systems, the scaling theory of localization predicts that no such transition occurs, and the localization length $\xi$ is always finite~\cite{ALscaling}. Despite its simplicity, the scaling theory of localization relies on several hypotheses, one of which is that, if one could switch off interference, the transport would be properly described by a diffusion equation, i.e. that the scattering potential can be described as white noise. Once correlations are introduced in the scattering potential the picture becomes much less clear, and it is possible to have frequency bands where the system is localized co-existing with frequency bands where all the eigenmodes are extended~\cite{ALcoexisting, stockmann}, discrete sets of extended modes in an otherwise localized spectrum~\cite{discreteextended}, enhanced localization~\cite{ALenhanced}, or even fully extended Bloch modes in random-like potentials~\cite{supersymmetry}.

In this article we study experimentally the case where the scattering potential is described by colored noise instead of white noise, i.e. when the power spectrum of the random potential is not flat. We show that, as the disorder becomes more \textit{colored}, the localization length becomes longer, until such point that it becomes significantly larger than the size of the scattering medium and the system does not show Anderson localization any more. The problem of Anderson localization in correlated potentials has been the focus of a lot of theoretical works (see e.g. Ref~\cite{ALcoexisting, ALaperiodic, ALselfaffine, ALshortrangecorr, ALLevy}), but so far there are only few experimental verifications~\cite{ALenhanced, stockmann}.

The case of a scattering potential characterized by a power-law spectral density $S(k) \propto k^{-\alpha}$, for some positive (and real) $\alpha$, was first discussed by de Moura and Lyra, who used renormalization group techniques to show that when $\alpha$ becomes larger than 2 a 1D system will show a band of unlocalized states~\cite{lyra}. This result can be understood if we consider that, for $\alpha \ge 1$, the spectral density $S(k)$ can not be normalized for an infinite system. For any system of finite length $L$ there is no problem, as the spectrum is effectively cut off for $k$ smaller than $\sim 1/L$, but this means that the proper normalization factor is now size-dependent, hence making $S$ (and thus the scattering properties) size-dependent. Nevertheless, it was predicted that for $1 \le \alpha <2$ the system is still localized, and the modes become extended only for $\alpha \ge 2$~\cite{lyra,roemer}, which is surprising if one considers that larger values of $\alpha$ correspond to a scattering potential ever closer to a sinusoid, and that a small amount of disorder in an otherwise perfectly periodic potential is well known to enhance Anderson localization~\cite{john}. Furthermore long-range correlations were recently associated with strengthened localization~\cite{ALenhanced}.

\begin{figure}[b]
	\centering
\includegraphics[width=0.4\textwidth]{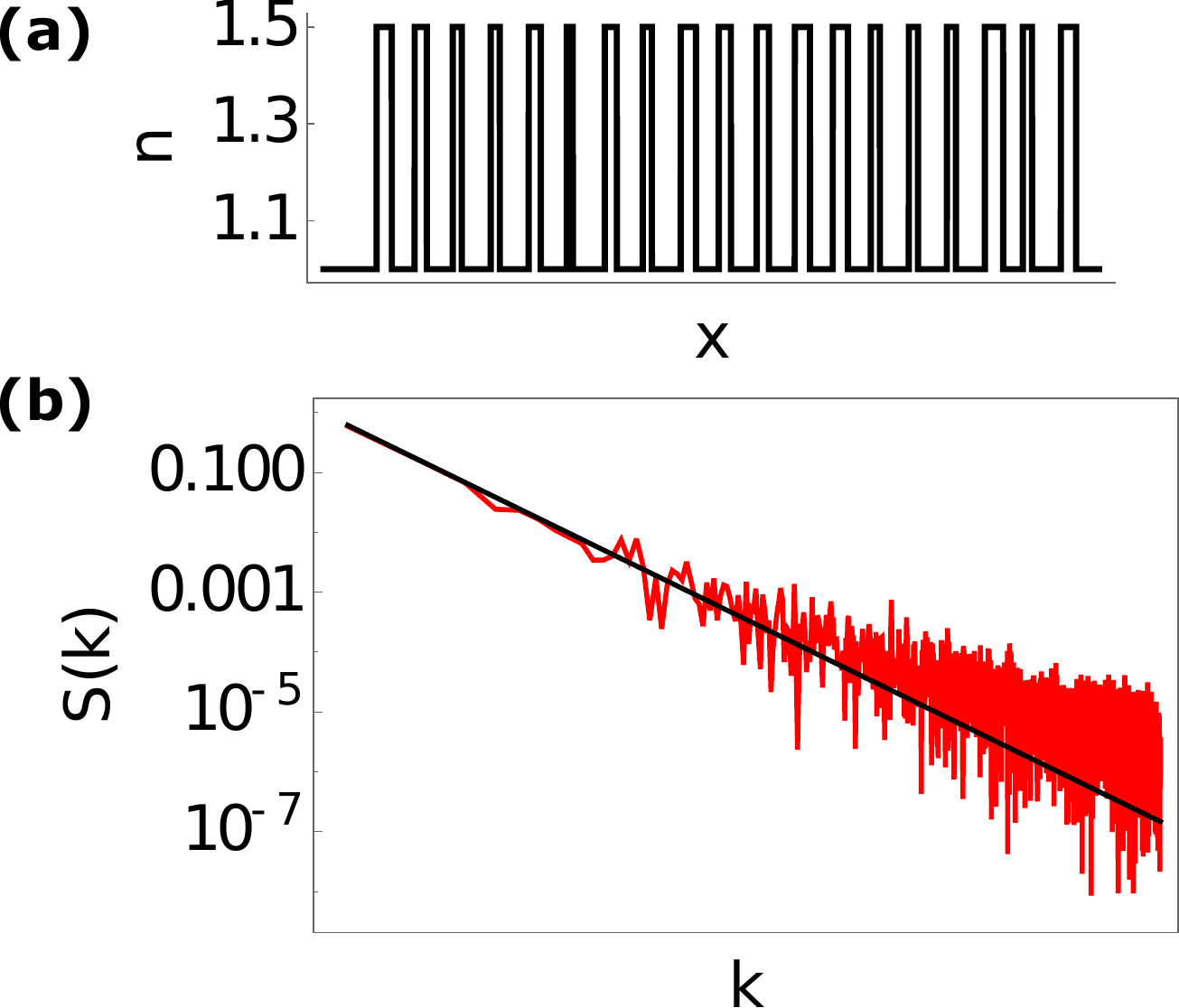}
	\caption{(a) Schematic of the scattering potential: a set of slabs with refractive index $n=1.594$ with different thicknesses are positioned at 2.5~cm intervals in the waveguide. The sequence of thicknesses $d'$ is obtained using eq.~\ref{eq:thicknesses} to discretize the random sequence $d$ (eq.~\ref{eq:sequence}). (b) The sequence $d$ is designed to be random, but also to have long-range correlations, in the form of a power-law spectral density $S(k) \propto k^{-\alpha}$. The black line shows an example for $\alpha=2$. The discretization process lead to a sequence that has the same power-law spectral density, albeit noisier (red line).}
	\label{fig:setup}
\end{figure}

As an experimental testbed we generated spatially varying 1D scattering potentials in the form of refractive index profiles by placing slabs of acrylic with variable thickness within a 3~m long WR90 aluminum waveguide. The waveguide has an operating bandwidth of 8.2 to 12.4~GHz (corresponding to a wavelength range of 2.4~cm to 3.7~cm), and a removable top to allow accurate placement of the acrylic scatterers. The waveguide was marked at 2.5~cm intervals with one acrylic slab placed at each position such that, for a purely periodic structure (i.e. if all the slabs have the same thickness), the system would exhibit a pass band centered at 10~GHz. A random sequence with the desired spectral density can be obtained, as described in ref.~\cite{lyra}, by first generating $N$ uniformly distributed random numbers $\phi_m$ in the range $[0,2 \pi]$, and then using them to compute the sequence $d$ with elements
 \begin{equation}
 d(j) = \sum_{m=1}^{N/2} \sqrt{m^{-\alpha} \left| \frac{2 \pi}{N} \right|^{1-\alpha}} \cos \left( \frac{2 \pi j m}{N} + \phi_m \right) .
 \label{eq:sequence}
 \end{equation}
 Since experimentally we can use only a discrete set of (positive) thicknesses, the sequence $d$ was shifted and rescaled (which does not change the shape of the power spectrum) as
\begin{equation}
d'(j) = \left \lceil 2.5 \left( \frac{d(j)}{\vee \left| d \right|} +1 \right) +1 \right \rceil,
\label{eq:thicknesses}
\end{equation}
where $\vee \left| d \right|$ is the maximum value of the absolute value of the sequence $d$, and $\lceil . \rceil$ is the ceiling function, which returns the smallest integer bigger than a given input (e.g. $\left \lceil 3.14 \right \rceil=4$). This produces a sequence of integer numbers between 2 and 6~mm. As shown in Fig.~\ref{fig:setup}, the power spectrum of the discretized sequence still has the desired power-law behavior, albeit noisier than the ideal one. It is worth noticing that in our experiment it is the sequence of the scatterers' thickness to follow explicitly the desired distribution, but that the position-dependent refractive index $n(x)$ must have the same power spectrum by construction.

\begin{figure}[b]
	\centering
\includegraphics[width=0.4\textwidth]{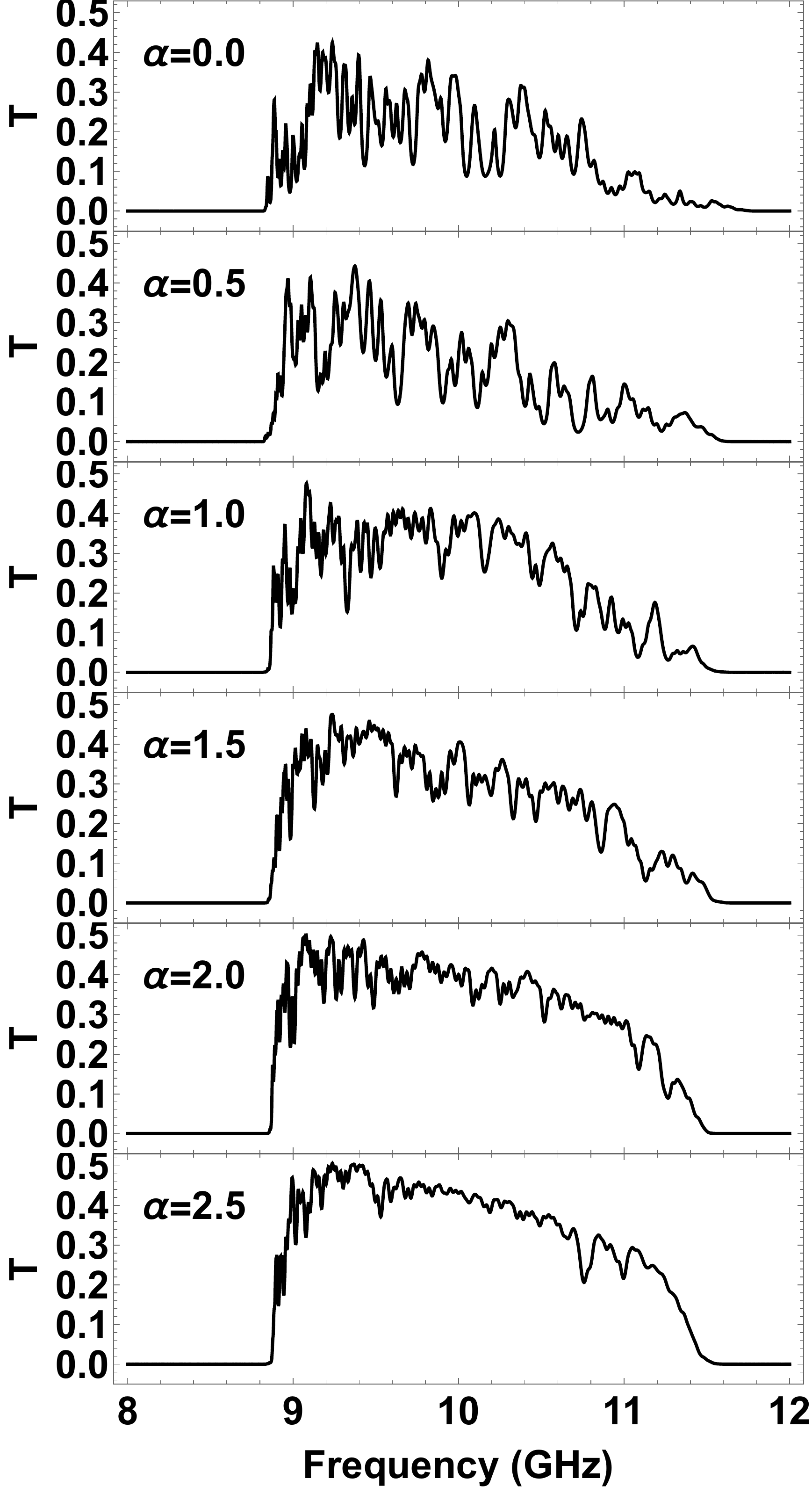}
	\caption{Typical transmission spectra for different values of $\alpha$. For low values of $\alpha$ the peaks corresponding to single Anderson localized modes are clearly visible, but as $\alpha$ increases the modes become wider and start to merge into each other, forming a smooth transmission band for large values of $\alpha$.}
	\label{fig:spectra}
\end{figure}

\begin{figure}[tb]
	\centering		
\includegraphics[width=0.4\textwidth]{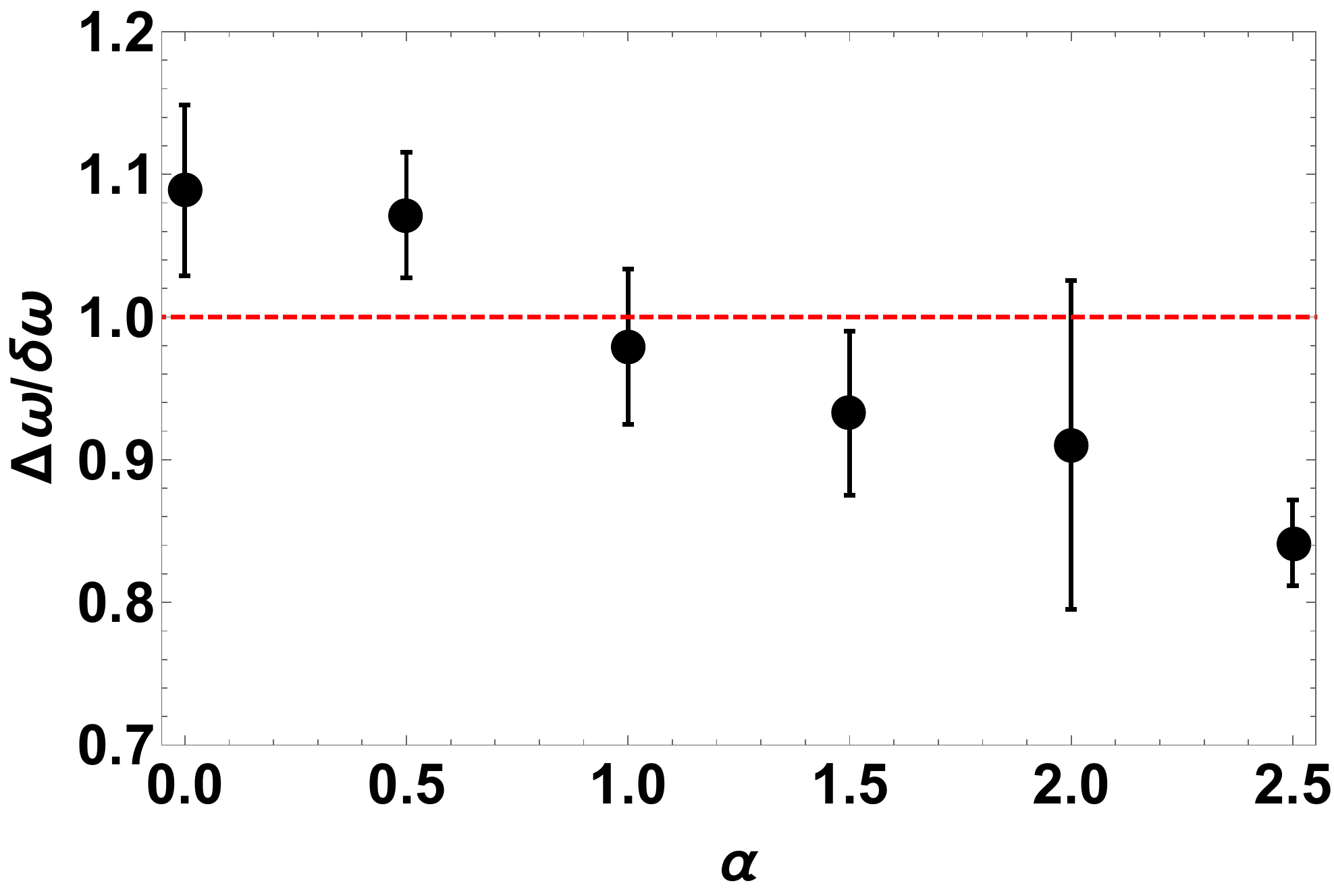}
	\caption{Ratio of the mode separation $\Delta\omega$ to the mode width $\delta\omega$ as a function of $\alpha$. Performing a multipeak fit of the transmission spectra we obtained the average spacing between the peaks and their average width. The error bars represent the standard deviation of the ratio $\Delta\omega / \delta\omega$ over the different realization of the disorder. The Thouless criterion for Anderson localization states that a 1D system can be considered localized if this ratio is larger than 1 (dashed red line), i.e. if the (localized) modes are spectrally separated. Once the ratio becomes smaller than 1 we can not talk about separated modes anymore and the system becomes conductive again.}
	\label{fig:modeseparation}
\end{figure}

After undertaking a standard Through-Reflect-Line calibration, the reflection from, and transmission through, the waveguide was measured using an Anritsu VectorStar Vector Network Analyser. We repeated the measurements for systems with $\alpha$ varying from 0 to 2.5 in steps of 0.5, and for 10 different realizations of the disorder for each value of $\alpha$. Fig.~\ref{fig:spectra} shows typical transmission spectra for each value of $\alpha$. For low values of $\alpha$ the transmission peaks due to Anderson localization~\cite{AL1d} are distinct and clearly visible, while for higher values of $\alpha$ the peaks merge with each other and the spectrum becomes smoother. While difficult to apply to higher dimensional systems, for 1D structures the Thouless criterion offers a convenient way to discriminate between localized and non-localized disordered systems: if the typical distance $\Delta \omega$ between the modes is larger than their typical width $\delta \omega$, then the system is localized. To test this criterion we performed a multi-peak fit to the transmission spectra and calculated the ratio between the average $\Delta \omega$ and the average $\delta \omega$ for each spectrum~\cite{supp}. Fig.~\ref{fig:modeseparation} shows that, consistent with the qualitative observation in Fig.~\ref{fig:spectra}, the system becomes less and less localized as $\alpha$ increases, until it becomes fully delocalized for a value of $\alpha$ between 2 and 2.5.

It is worth noting that our experimental system, like all microwave and optical systems, suffers from losses. One small source of loss is the absorption that arises from the scatterers themselves, with a real part of their refractive index of $n=1.594 \pm 0.003$, and an imaginary part of $<0.005$~\cite{supp}, which is negligible with respect to the other loss channels. Other sources include the inherent losses in the metal of the waveguide (approximately 0.2~dB/m), and those arising from the small, but unavoidable, gap between the guide and its removable lid ($\sim 50~\mu m$). Therefore, we expect the resonances due to Anderson localization to have a lower Q-factor and thus be wider then expected for an ideal lossless system. As this can increase $\delta \omega$, it is necessary to double-check our results with a technique less susceptible to absorption and losses. One of the archetypal properties that distinguish a diffusive system from an Anderson localized system is how the total transmission decreases with the sample thickness: linearly for a diffusive system, and exponentially for a localized one~\cite{ALwiersma}. As absorption also leads to an exponential decay of the transmission with thickness it is of fundamental importance to distinguish the two effects~\cite{ALwiersma,ALsheffold,ALwiersma2}.
In a 1D geometry, such as the one in our experiment, absorption can be measured independently from localization by measuring both the total transmission $T$ and the total reflection $R$ and computing the absorption as $A=1- (T+ R)$. Once $A$ is carefully characterized, we can determine how much of the exponential decrease of $T$ with thickness can not be explained by absorption, and use this to estimate the localization length. To do so we measured $T$ and $R$ for samples containing between 1 and 118 scatterers, i.e. samples with total length $L$ between 2.5 and 295~cm long, thus obtaining an $A$ vs $L$ curve for each value of $\alpha$. As the empty waveguide has finite losses, we expect that $A(0)=A_0 >0$. Furthermore, for large $L$, we expect the contribution of the transmitted intensity to $T+R$ to be negligible, and the contribution of the reflected intensity to saturate to some value $R_0$, i.e. making the sample increasingly long will not lead to a higher value of R since absorption will prevent the fields that have penetrated far into the sample from ever being reflected.
\begin{figure}[tb]
	\centering		
\includegraphics[width=0.4\textwidth]{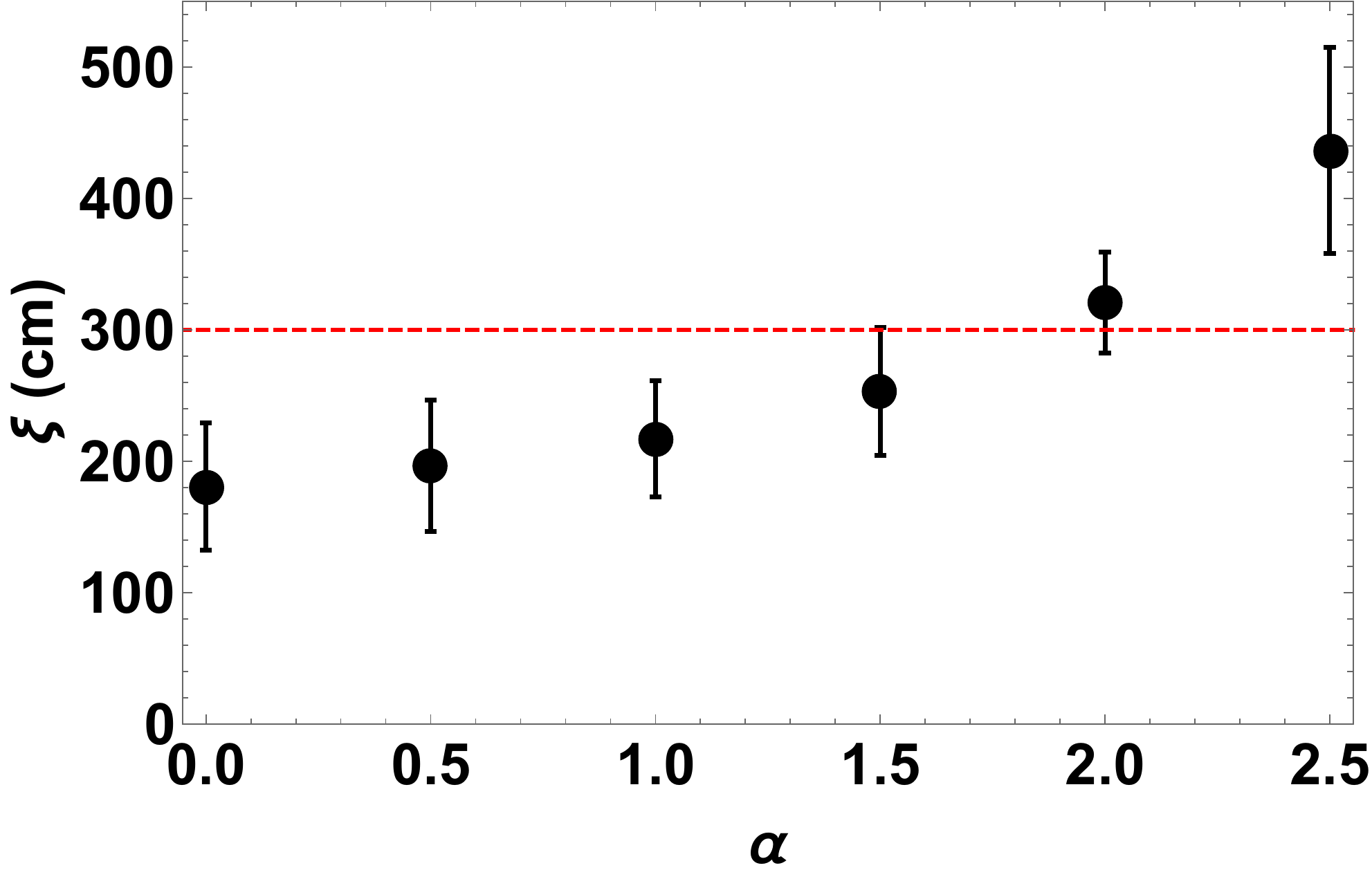}
	\caption{Values of the localization length $\xi$ as a function of $\alpha$. The error bars represent the fit uncertainties for $\mu$ and $b$ propagated to $\xi$. The part of the exponential decay of $T(L)$ that can not be accounted for by absorption gives us an estimate of the localization length $\xi$. For low values of $\alpha$ the measured localization length is significantly shorter than the total sample thickness (shown by the red dashed line), but when $\alpha \gtrsim 2$ there is a cross-over and the system stops being localized.}
	\label{fig:xi}
\end{figure}
Therefore we can fit our absorption curves to 
\begin{equation*}
A(L) = (1- R_0) + e^{- \mu L} (A_0+ R_0 -1),
\end{equation*}
where $L$ is the sample length, and $\mu$ the absorption coefficient. For each value of $\alpha$ we repeat the experiment 10 times to average over the realizations of disorder, and obtain a value for $\mu$. If no localization was occurring, a fit to $T(L)$ with the exponential function $T(L)= T_0 e^{- b L}$, where $T_0$ is the transmission coefficient of the empty waveguide, would yield $\mu=b$ within the experimental error. As we obtain $b > \mu$ for all the values of $\alpha$ and all the realizations of disorder, we can deduce that absorption can not explain the exponential decay of $T(L)$ alone. Therefore we interpret $\xi = \frac{2}{b-\mu}$ as the localization length of the system~\cite{supp}. Fig.~\ref{fig:xi} shows the obtained values of $\xi$ as a function of $\alpha$. This shows that for values of $\alpha$ greater than approximately 2, the localization length exceeds the sample size. This is consistent with the results in Fig.~\ref{fig:modeseparation}.
It is important to note that the localization length we measure is averaged over the whole spectral range, and that some frequency bands delocalize faster than others. As can be seen in Fig.~\ref{fig:spectra}, the modes at the centre of the band (in our case around 10~GHz) delocalize faster than the modes at the band edge~\cite{lyra}.

In conclusion, we have experimentally shown that correlations in the disordered potential can make a 1D scattering system non-localized. In particular we have shown that this happens when the scattering potential has a power-law spectral density with a power $\ge2$. As previous results have shown that other kinds of long-range correlations can lead to an enhanced Anderson localization \cite{ALenhanced} (i.e. exactly the opposite effect), this proves that one needs to be very careful when talking generically about short or long-range correlations, as different flavors of correlations can give very different results.

\begin{acknowledgments}
We wish to thank R.A. R\"{o}mer for stimulating discussions.
\end{acknowledgments}

\section*{Appendix A: The experimental apparatus}
The WR90 aluminum waveguide was machined to have an operating bandwidth between 8.2 and 12.4~GHz. The complex reflection and transmission amplitude coefficients from/through the waveguide were measured using an Anritsu VectorStar Ms4640B Vector Network Analyser (VNA) after a standard Through/Reflect/Line waveguide calibration~\cite{calibration}. The scattering slabs were laser-cut from extruded acrylic sheets of different thickness. There was a small variation in the thickness of each acrylic sheet across its surface resulting in a spread of thickness for each scatterer type. The mean and standard deviation for each type of scatterer was:

\begin{center}
 \begin{tabular}{|c|c c |} 
 \hline
Scatterer nominal  & \multicolumn{2}{|c|}{Measured thickness:} \\ 
thickness (mm) & Mean (mm) & $\sigma$ (mm)\\
 \hline\hline
2  &2.05 & 0.01 \\ 
 \hline
3  &2.89 & 0.02 \\
 \hline
4  &3.74 & 0.01 \\
 \hline
5  &5.07 & 0.01 \\
 \hline
6  &5.63 & 0.01 \\  
 \hline
\end{tabular}
\end{center}

In order to determine the complex refractive index of the acrylic slabs we measured the complex reflection and transmission amplitude coefficients from/through a single 5.63~mm thick slab within the waveguide and extracted the permittivity and permeability of the acrylic using the Nicholson-Ross-Weir algorithm~\cite{NRW1,NRW2}, resulting in a real part of the refractive index of $\Re(n)=1.594 \pm 0.003$. The imaginary part of the refractive index was too small to be accurately measured by this method ($<0.005$), but the losses in the acrylic are much smaller than those within the metal of the waveguide and can therefore be neglected.\\

\section*{Appendix B: Multipeak fit}

\begin{figure}[bht]
	\centering		
\includegraphics[width=0.4\textwidth]{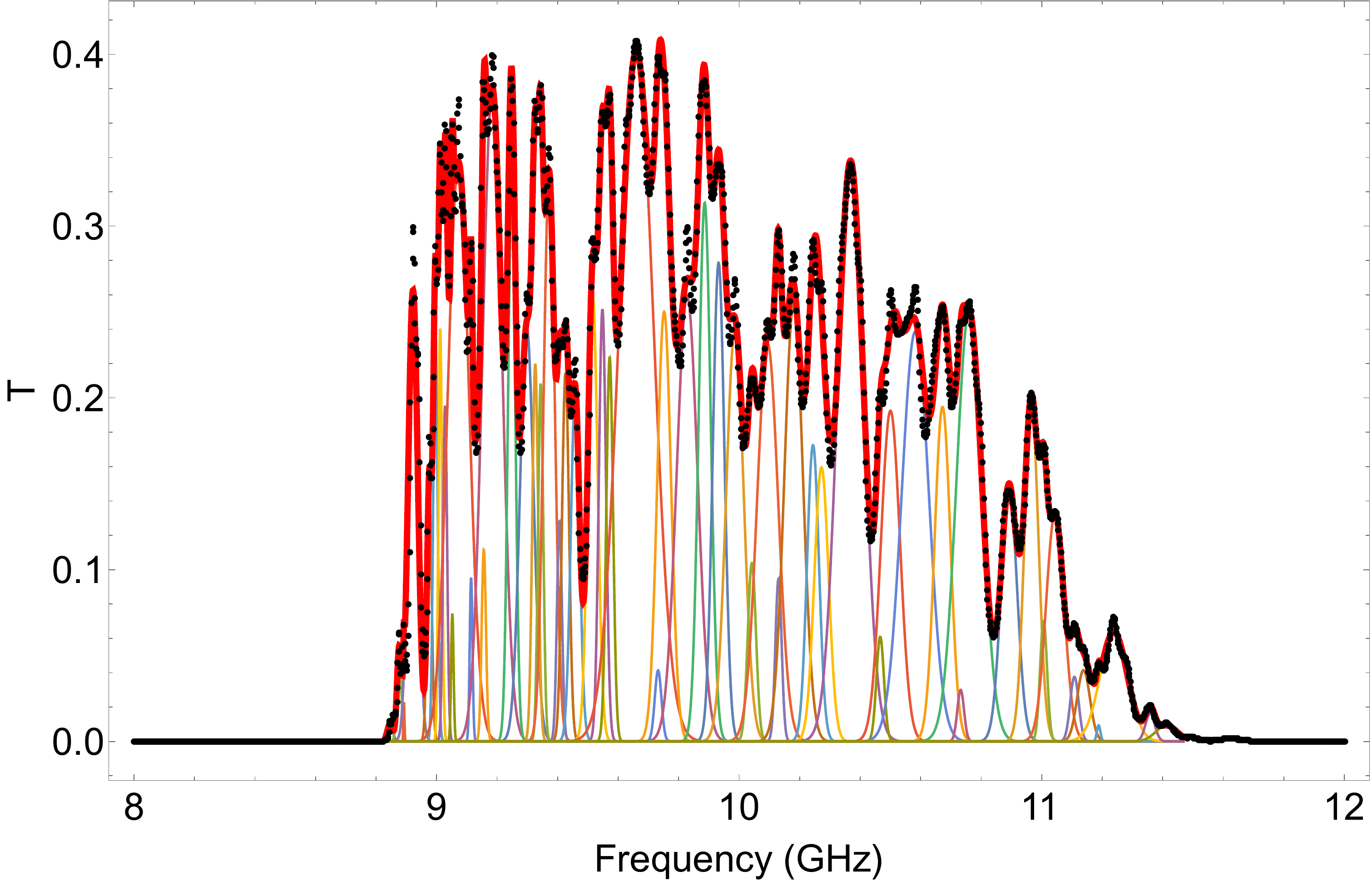}
	\caption{Example of a multipeak fit for a transmission spectrum at $\alpha=0.5$. The black dots are the experimental data points, colored thin lines are the fitted peaks, and the thick red line is their sum (which has to be compared to the experimental data).}
	\label{fig:fit}
\end{figure}

For each value of $\alpha$ we selected the 10 transmission spectra (corresponding to the 10 realizations of the disorder) for the longest sample, i.e. 118 scatterers. For each spectrum we first estimated the number of peaks by looking at the local maxima, and then fitted an equal number of Gaussian functions to the spectrum, using the peaks' height and width as fit parameters, while their position was kept fixed to the point of the local maxima. Figure~\ref{fig:fit} shows an example for $\alpha=0.5$. Shown are the the data points, the 55 Gaussian functions used to fit it, and their sum. From this fit we can extract the average distance between neighboring peaks $\Delta\omega$, and the average peak width $\delta\omega$. Repeating this for all 10 realizations of the disorder we can then estimate the variance for these two values.\\
It is important to notice that this method to fit the transmission spectra is bound to underestimate the number of Gaussian functions needed, as it only considers the modes that form a local maximum in the spectrum. Therefore both $\Delta\omega$ and $\delta\omega$ are overestimated. This is confirmed by the fact that the $\chi^2$ for each fit is always significantly larger than the number of degrees of freedom, which would be the expected value for an ideal fit. This introduces a systematic error that is difficult to estimate. Apart from this systematic error the results obtained are both numerically stable and repeatable, and the qualitative picture that emerges is consistent with theory~\cite{lyra}. An alternative approach would be to use as many modes as there are scatterers (which is roughly the number of modes we expect to contribute for a 1D system~\cite{AL1d}), but any fit with over 300 free parameters is bound to be numerically unstable, and thus would not provide more reliable results than the method we used here.

\section*{Appendix C: Estimate of $\xi$}
\begin{figure*}[th!]
	\centering		
\includegraphics[width=0.8\textwidth]{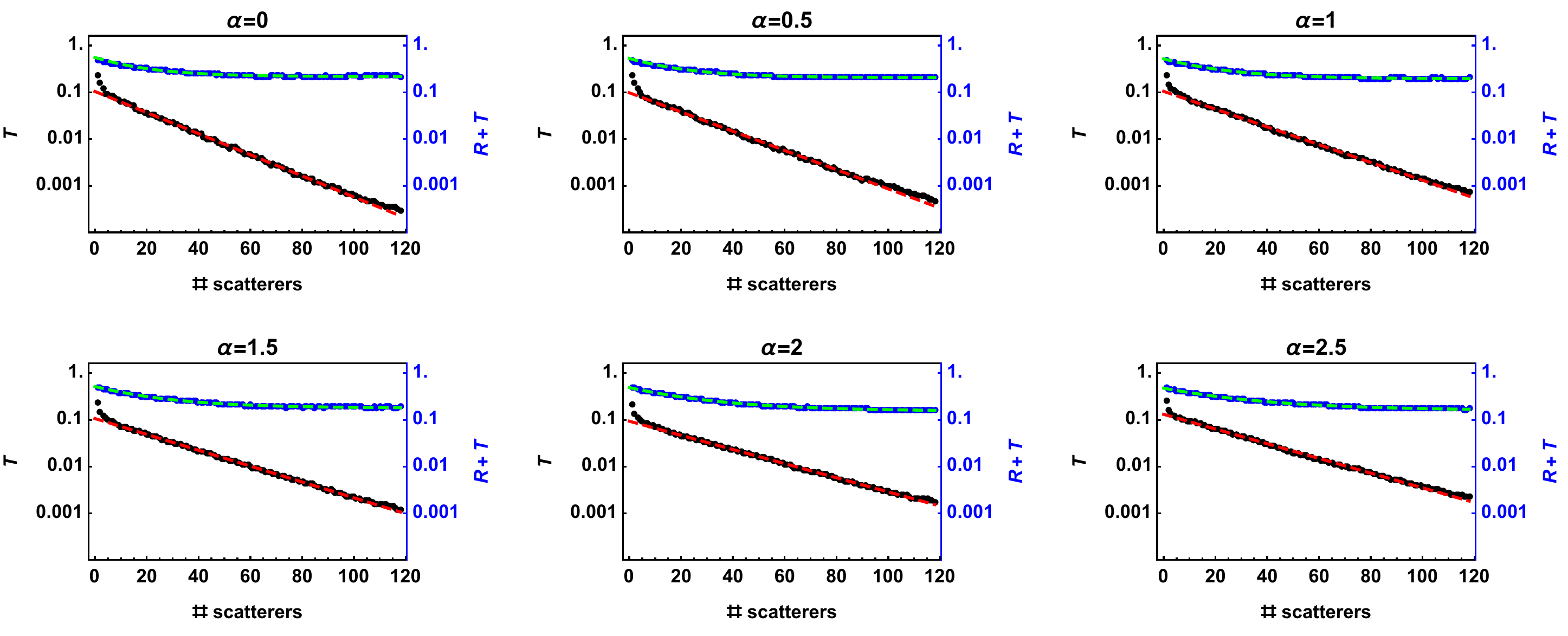}
	\caption{Fits of $T(L)$ (black points) and $T(L)+R(L)$ (blue points) to obtain, respectively, $b$ and $\mu$, for the various vales of $\alpha$.}
	\label{fig:xi}
\end{figure*}
For the reasons explained above, the multipeak fit of $T(\omega)$ provides a qualitative description of the transition between an Anderson localized regime to a conductive one when increasing $\alpha$, but not a quantitative one. To measure directly the (spectrally averaged) localization length $\xi$ we look at how the total transmission $T$, and the sum between the total transmission and total reflection $T+R$, change with the length of the sample. For a non absorbing system $T+R$ should always be equal to 1 independently from the transport regime, so any change can be attributed to losses. If we assume that the losses are uniform along the waveguide, we can fit $T+R$ to $(T+R)(L)=1-A(L) = 1- \left[ (1- R_0) + e^{- \mu L} (A_0+ R_0 -1) \right]$, where $A_0$ is the total absorption of the empty waveguide and $\mu$ the absorption coefficient. In order to do an ensemble average we averaged both $T(L)$ and $(T+R)(L)$ over the 10 realizations we measured, and perform the fits, as shown in Fig.~\ref{fig:xi}.

To estimate $\xi$ we look at how much of the exponential decay of $T(L)$, parametrised as $T(L)= T_0 e^{- b L}$, can be ascribed to losses, and how much can not. In a perfectly conductive regime losses are the only explanation for the exponential decay of $T$ with the sample length, but if the system is Anderson localized $T$ will decay faster than is predicted by the losses. Therefore we estimate the localization length as $\xi = \frac{2}{b-\mu}$. The variance on the parameter estimation on $\mu$ and $b$ (in the least square fitting) was used to obtain an uncertainty on the value of $\xi$.

\end{document}